\documentclass[%
 reprint,
 amsmath,amssymb,
 aps,
]{revtex4-2}

\usepackage{dcolumn}
\usepackage{bm}

\usepackage{graphicx}
\usepackage{caption}
\usepackage{subcaption}
\usepackage{graphicx,hyperref}
\expandafter\let\csname equation*\endcsname\relax
\expandafter\let\csname endequation*\endcsname\relax
\usepackage{amsmath}
\usepackage{amssymb}
\usepackage{multirow, makecell, comment} 
\newcommand{\fla}[1]{\begin{flalign}#1\end{flalign}}
\usepackage{braket}
\usepackage{lineno}
\usepackage{lipsum}
\usepackage{xcolor}
\usepackage{soul}
\usepackage{array}

\begin{document}

\preprint{APS/123-QED}
\preprint{APS/123-QED}
\title{Dynamic control of resonance fluorescence in graphene quantum plasmonics.}


\author{Ali.A. Kamli$^1$}
\author{S.A.Moiseev$^2$}
\author{Jabir W. Hakami$^1$}
\affiliation{$^{1}$ Department of Physics, Jazan University, Jazan Box 114, Saudi Arabia.}
\affiliation{$^{2}$Kazan Quantum Center, Kazan National Research Technical University n.a. A.N. Tupolev-KAI, 10 K. Marx St., 420111, Kazan, Russia,
\\
 E-mails: 
 aamk0083@gmail.com,
 s.a.moiseev@kazanqc.org}

\date{\today}

\begin{abstract}

The spectral and statistical properties are explored for surface plasmon (SP) emission in resonance fluorescence from a driven two level emitter in the proximity of 2D single graphene sheet. We derive an exact closed form analytic expression for the emitted SP field valid in the near and far regions. The SP field profile and spectrum function depend on the graphene conductivity and take into account the dynamic control parameters, namely Fermi energy. We present analysis for the spectrum and second order coherence functions and discuss the possibility of their control using graphene system parameters to manipulate the spectral linewidth and second order coherence function.  
\end{abstract}

\maketitle

\section{Introduction}

Control and manipulation of radiation from quantum emitters is of great interest at fundamental and practical level. Resonance fluorescence (RF) and its emitted radiation are intriguing problems in quantum physics that have for many decades fascinated researchers and scientists with their counter intuitive features\cite{Kimble1976,Kimble1977,Mandel1995,Scully1997,Carmichael2002}. The problem one is usually trying to address is that of a two level emitter (TLE) resonantly driven by a classical monochromatic laser field where it jumps from its ground to excited state. Once in the excited state the emitter relaxes to the ground state again by the process of spontaneous emission. The fluorescence spectrum of emitted radiation and its statistical properties are measured. The behavior of spectrum depends strongly on the relative size of the driving field and the TLE linewidth. For driving field Rabi frequency smaller than the linewidth, the spectrum shows a single peak centered at the resonant frequency. In the strong field limit when the Rabi frequency is larger than the linewidth the spectrum splits into the main single peak in addition to two side band peaks first predicted by Mollow \cite{Mollow1969}, that are attributed to dynamic Stark coupling due to the strong coupling of TLE with the field and may be explained in terms of the dressed state picture\cite{Scully1997}. Furthermore, the emitted radiation characterized by the coherence functions, showing non-classical light features namely anti-bunching and squeezing effects \cite{Carmichael2002}. These RF interesting features have been explored long ago theoretically \cite{Knight1980,Loudon1980} and experimentally \cite{Schuda1974,Wu1975}. Resonance fluorescence has also been extended to three and and multi-level systems \cite{Carreno2017,Kosionis2022} in various physical schemes including superconducting quantum dots in planar cavity \cite{Muller2007}, solid state molecules \cite{Wrigge2007}, artificial atoms \cite{Astafiev2010} squeezed vacuum \cite{Toyli2016}, near a plasmonic structure \cite{Klimov2012}, a quantum dot plasmonic structure\cite{Yang2016} and in V-type emitter coupled to metal nanoparticle \cite{Kosionis2022}. Recently resonance fluorescence has been suggested as a source of entangled photons utilizing Mollow triplets to entangle photons emitted from the sidebands \cite{Carreno2024}. More recently there have been renewed interests both theoretically and experimentally to explain the origin of coherence and antibunching, and the spectral behavior of radiation emitted in resonance fluorescence \cite{Carreno2018,Hanschke2020,Wang2025}. 

In this work we are interested to explore spectrum and coherence properties of a two level emitter coupled to the surface plasmon polariton modes in proximity of a 2D graphene thin sheet of conductivity $\sigma$ immersed between two much thicker semi-infinite media of dielectric media. Surface plasmons are known \cite{Barnes03,Zayats05,Maier07,Tame2013} to couple strongly to dipole emitters. On the other hand, the pioneering work of Purcell \cite{Purcell1946,Berman1994,Milonni94} has established the fact that the emitted radiation linewidth is influenced by the environment in which the emitter is located. It is therefore legitimate to ask how the spectral and statistical properties of the emitted radiation are affected by plasmonic environment that surrounds the driven two level emitter. Surface plasmon polaritons (SPs) are the electromagnetic modes of the dressed light-matter coupling that takes place at the interface of metal like material media. These SPs have long enjoyed interesting optical properties that made them the focus of intense research. An important property of SPs is their ability to confine light into nanosclae regions of space due to the highly confined nature of the fields near interfaces.
The SPs field confinement has always been associated with energy losses in media, and besides tunability this has been a major obstacle in plasmonics  \cite{Kamli08,Moiseev10,Ozbay06,Gram2010}. 

Recently graphene plasmonics has been suggested as a new platform for plasmonics \cite{Geim2011,Novoselov2011,Novoselov2005,Falkovsky2008,Koppens2011,Abajo2014,Goncalves2016}. 
Graphene plasmonics are more tunable due to graphene chemical doping or electric gating, with its conductivity can be varied with many parameters like Fermi energy, temperature, field mode, and electron density so that graphene surface plasmons (GSP) can be well tuned to desired frequency using such varied parameters. Their propagation wave number parallel to graphene interface is much larger than the free space wave number leading to evanescent decay on both sides of graphene interface and thus highly confined fields.   
The GSP wavelength is thus much shorter than the free space wave length which can confine light into nanoscale regions with  fields confined to very small volumes much smaller than the diffraction limit. Due to this mismatch of energy (in terms of frequency) and momentum (in terms of wave number) direct excitation of GSPs is not possible. Special excitation techniques are used to fulfill conservation of energy and momentum, such techniques include grating coupling, prism coupling, highly focused optical beams and near-field excitation techniques to mention just few examples \cite{Maier07,Goncalves2016,Novotny2012}. The strong coupling of GSPs to light results in huge optical enhancements which means that due to their strong interaction with light, the resultant SP field near interface has larger intensities than the exciting fields,
which have already opened up exciting possibilities for applications in the field of surface enhanced Raman scattering (SERS) \cite{Ru2008}.

Furtheremore GSPs enjoy low losses below the Fermi energy which means they have longer propagation distance, which can be 10 and more $\mu$m, that opens up unique opportunities for the realization of coherent strong interaction of plasmons with atoms. An outstanding property of graphene is its high absorption coefficient, which for undoped suspended graphene in air is about $\pi\alpha\approx2.3\%$ independent of wavelength in the visible range of spectrum, where $\alpha=1/137$ is the atomic fine structure constant \cite{Koppens2011,Goncalves2016}. In resonant structures this value of absorption coefficient can be higher and may be frequency dependent.    
These exciting properties make GSP attractive for exploring the interaction of GSP with quantum emitters at the fundamental level and their applications in quantum technologies 
\cite{Mikhailov2007,Huidobro2012,Nikitin2011,Forati2014,
Huang2017,Thanopulos2022,Ferreira2020,Grigorenko2012,cui2021graphene,Eriksen2025}.

Our goal is therefore to explore the various effects of the graphene plasmonics on the spectral and statistical properties of the flourescence radiation emitted by the driven two level emitter near a graphene 2D sheet namely the effects of transition dipole orientation angles, distance from the graphene sheet and Fermi energy etc. We shall pay special attention to the role of Fermi energy as a dynamic control to explore its effect in resonance fluorescence calculations. 
We shall derive an exact closed analytic expression for the SP emitted field that is valid in the near and far field case and analyze its profile for various graphene structure parameters and dipole orientations. This is particularly interesting and important because in the usual treatment of resonance fluorescence one is usually interested in the far-field case where the radiation field is given by its approximate expression in the far regions of space and ignoring near and intermediate field contributions. Such contributions are particularly important in the surface plasmon polaritons since by their very nature they are near fields and one must retain full expression for the radiation field not only the far field part. This major modification is likely to introduce significant changes in the emitted spectrum and its statistical properties which we like to explore here. The analysis of resonance fluorescence adapted here for GSP is very important and potentially useful for integrated quantum photonics.     

\begin{figure}   
\centering
    {\includegraphics[width=8cm]{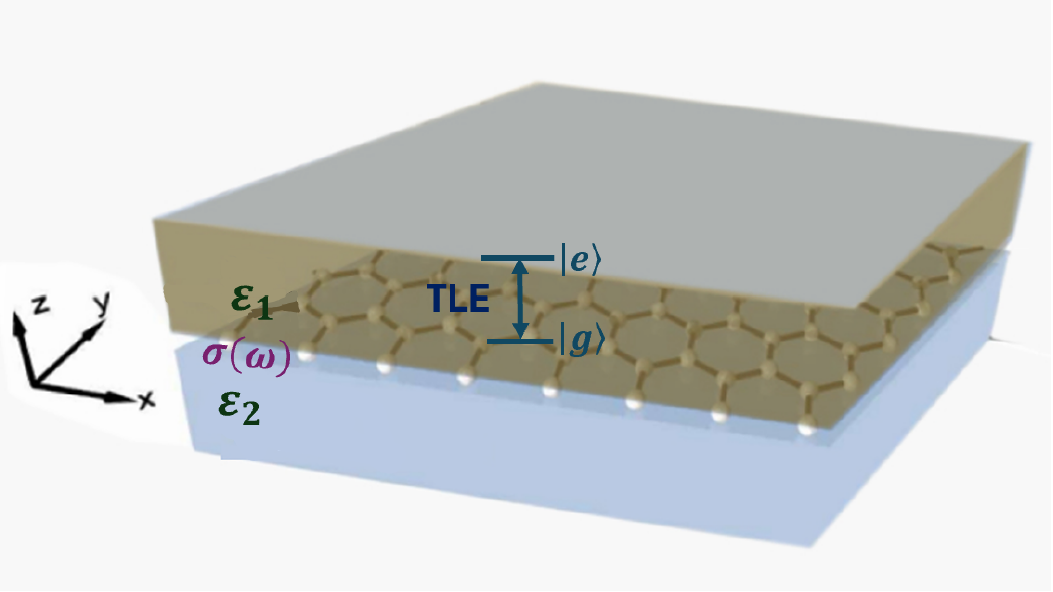} }
    \caption{2D single graphene sheet of conductivity $\sigma(\omega)$ between two semi-infinite half spaces with constant dielectric functions $\varepsilon_1$ and $\varepsilon_2$. A two level emitter is placed above graphene sheet in upper medium. }
    \label{fig:fig1}
\end{figure}

\section{Graphene plasmonics environment}

 Fig.1 shows the simplest system of graphene based plasmonics.  It consists of a 2D thin layer of graphene of surface conductivity   $\sigma$ positioned at interface  $z=0$ and sandwiched between two much thicker dielectric layers occupying the upper $z>0$ and lower $z<0$ half spaces. The two dielectrics are taken to have constant real dielectric functions $\varepsilon_1$ and $\varepsilon_2$. The 2D graphene sheet  supports both transverse magnetic (TM) and transverse electric (TE) \cite{Mikhailov2007} polarized surface plasmon modes propagating along in-plane unit vector $\hat{u}_{||}$ such that the SP electric field amplitudes decay away in both sides with distance from the interface at $z = 0$ with high energy concentration close to interface. The surface plasmon modes are thus solutions to the homogeneous wave equation and for TM polarized modes, the transverse fields $\mathbf{\textit{E}}$ of frequency $\omega$ satisfy wave equations and have the general forms
\fla{
{\textbf{E}}(\textbf{r},k_{||})=& e^{iK_{||}.r_{||}}
[u(z) 
(\hat{\textbf{u}}_{||}+\hat{\textbf{z}}\frac{i K_{||}}{\beta_1})e^{-\beta_1 z} A 
\nonumber \\
+& u(-z) 
(\hat{\textbf{u}}_{||}-\hat{\textbf{z}}\frac{iK_{||}}{\beta_2}) e^{\beta_2 z} B ],
\label{Fields-1}
}

\noindent
where u(z) is the step function. The complex wave numbers parallel and normal to interface $K_{||}=k_{||}+i\kappa$ and $\beta_j$ are related as 
$\beta_{j}=(K_{||}^2-\varepsilon_j k_0^2)^{1/2}$, $k_0=\omega/c$ for j=1,2, where $k_{||}$ and $\kappa$ are real and imaginary parts of $K_{||}$. Both wave numbers $K_{||}(\omega), \beta_j(\omega)$ are functions of mode frequency $\omega$, but we suppress this dependence for convenience.  
The constants $A$ and $B$ are to be determined from interface boundary conditions and SP field quantization shortly.
The TE polarized SP modes electric fields can be written in a similar form. 
The presence of the graphene two-dimensional charge sheets at the  interface presents jump conditions on the tangential component of the magnetic fields in addition to the continuity of the tangential component of the electric field. The magnetic fields associated with these electric fields are calculated from the relation $i\mu_0\omega\textbf{H}=\bigtriangledown\times\textbf{E}$, where $\mu_0$ is vacuum permeability. These two conditions on continuity of the tangential components of the electric fields and jump condition on tangential components of the magnetic fields at z=0 lead to 
\fla{
&\textbf{E}_1^{||}=\textbf{E}_2^{||},
\nonumber
 }
\fla{
&\hat{z}\times[\textbf{H}_{1}-\textbf{H}_{2}]=\sigma\textbf{E}_1^{||} ,
}
\noindent
where $\sigma$  is the graphene two dimensional surface conductivity, given below. Implementing the above boundary conditions leads to the following conditions for SP excitation dispersion as
\fla{
\frac{\varepsilon_1}{\beta_1}+\frac{\varepsilon_2}{\beta_2}+i\frac{\sigma}{\varepsilon_0\omega}=0 ,  (\text{TM modes})
\nonumber
\\
\beta_1+\beta_2-i\frac{\omega}{c}\frac{\sigma}{\varepsilon_0 c}=0,    (\text{TE modes})
\label{dispersion}
}

 We see from the above equations, that the effect of the 2D graphene sheet is quantified by its surface conductivity, which for the graphene sheet of interest here is given by the expressions \cite{Falkovsky2008,Koppens2011,Goncalves2016}:
\fla{
\sigma(\omega)=& \frac{e^2}{4\pi\hbar} \textit{f}(\omega,T,E_F),
\nonumber
\\
\textit{f}(\omega,T,E_F) = & 
\frac{i 8K_{B}T}{\hbar(\omega+i\tau^{-1})} \ln \left(2\text{cosh}[\frac{E_F}{2K_{B}T}]\right)
\nonumber
\\
+
[ \frac{\pi}{2}+\text{arctan}(\frac{\hbar\omega-2E_F}{2K_{B}T}) -&\frac{i}{2}\ln\frac{(\hbar\omega+2E_F)^2}{(\hbar\omega-2E_F)^2+(2K_{B}T)^2}].
\label{conductivity} 
}

The first term in this equation is the intra band contribution to conductivity coming from the electron transitions within the same band, and the remaining terms are the interband part from processes that involve electron transitions across the bands \cite{Goncalves2016}. The graphene conductivity is a temperature T dependent, it also depends on the Fermi energy $E_F$, and $\tau$. $K_B$ and $\hbar$ are the Boltzmann and Planck constants, $e$ is the electronic charge, $\tau^{-1}$ is electron scattering rate usually in the range $\tau\approx 0.01-1$ ps \cite{Koppens2011} and $\omega$ is the excitation frequency. Throughout this paper we shall assume room temperature $K_B T=0.03$ eV and take $\hbar\tau^{-1}=0.01$ eV.

\begin{figure}
    \centering
    \subfloat {(a){\includegraphics[width=8cm]{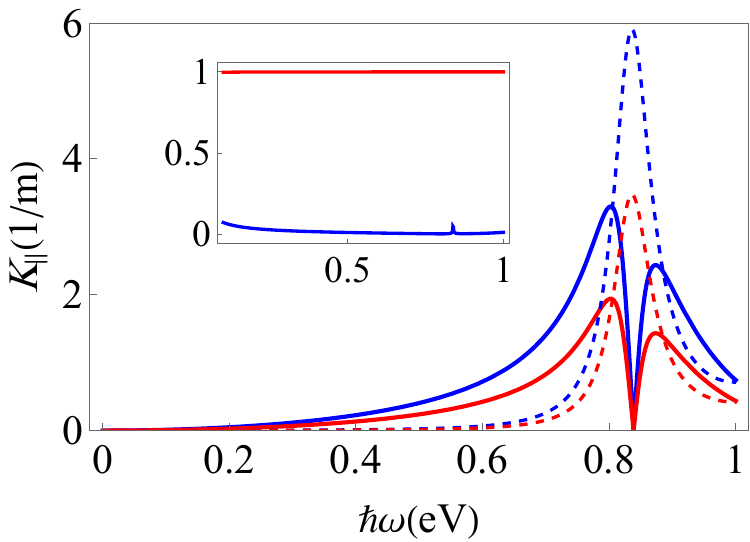}}}
    \qquad
    \subfloat{(b){\includegraphics[width=8cm]{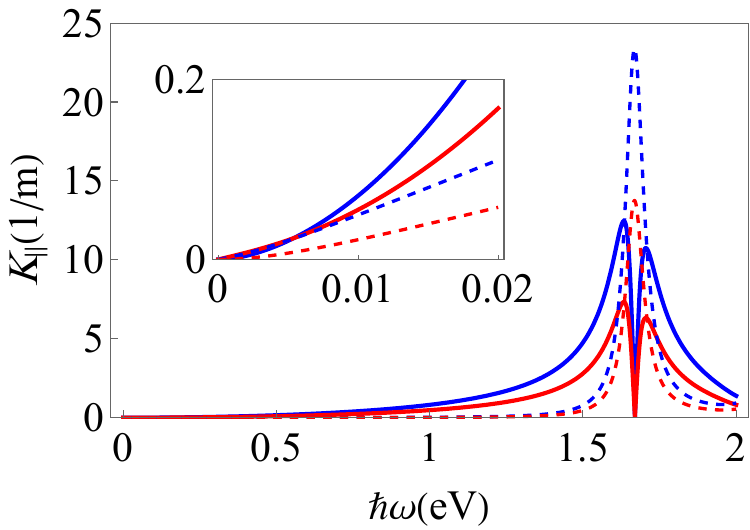}}}
    \qquad
    \subfloat{(c){\includegraphics[width=8cm]{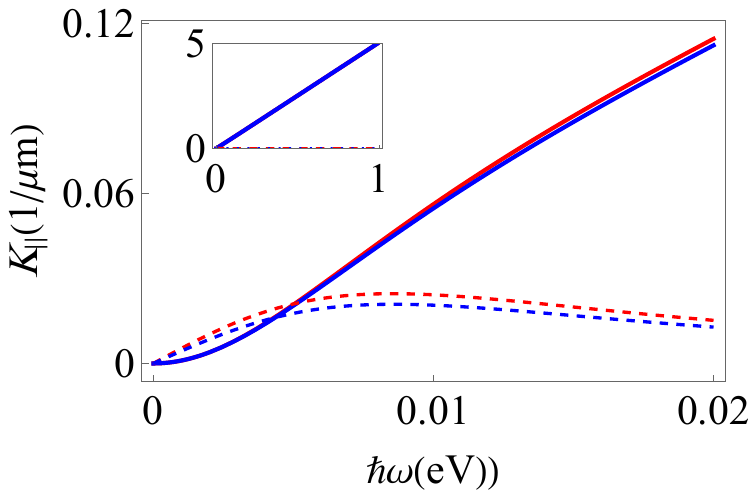}}}
    \caption{ Real and imaginary parts of graphene plasmonics complex wave number as functions of mode energy  for TM modes (a,b) and TE modes (c). In (a,b) the solid upper blue line shows $k_{||}=\text{Re}[K_{||}]$ for $\varepsilon_1=1,\varepsilon_2=2$, while solid lower red line gives $\varepsilon_1= \varepsilon_2=1$ case. The lower dotted lines are the corresponding imaginary parts. Fermi energy $E_F=0.5$ eV (a) and $E_F=1$ eV (b). The inset in (a) shows the group velocity $v_g/c$ for both TE (upper red line) and TM ( lower blue line).  The inst in (b) shows zoom in portion which displays losses at low frequency. In (c) the solid upper red line shows $k_{||}=\text{Re}[K_{||}]$ for $\varepsilon_1=1,\varepsilon_2=2$ case, while the solid lower blue line is for the $\varepsilon_1= \varepsilon_2=1$ case. The lower dotted lines are the corresponding imaginary parts. The inset in (c) shows an extended energy scale.}
    \label{fig:dispersion}
\end{figure}

The dispersions given by Eq.\eqref{dispersion} are shown in Fig.\ref{fig:dispersion} where we show the real $k_{||}(\omega)$ and imaginary $\kappa(\omega)$ parts of complex wave number $K_{||}(\omega)$ as functions of SP mode frequency $\omega$ for both TM (a,b) and TE modes (c) with graphene conductivity as in Eq.\eqref{conductivity} for $E_F=0.5$ eV and 1 eV. The wave number shows nonlinear dependence on frequency. To understand this behavior we consider the TM mode dispersion in the symmetric case i.e. $\varepsilon_1=\varepsilon_2=1$ which can be cast in the form $K_{||}(\omega)=\omega/c \sqrt{1-(2/\alpha f(\omega))^2}$, where
$\alpha=\frac{e^2}{4\pi\varepsilon_0c\hbar}=\frac{1}{137}$, is the atomic fine structure constant, the function $\textit{f}(\omega)$ is defined in Eq.\eqref{conductivity}. The wave number starts from zero for low frequency and increases and then suffers a dip when the quantity under the square root is zero which for the chosen parameters occurs at $\hbar\omega=0.837$ eV for $E_F=0.5$ eV. Then it increases again and drops again near Fermi energy. 
Here we point out a number of observations. 
From Fig.\ref{fig:dispersion} we note the wide range of tunability in the operating mode energy $\hbar\omega<0.8$ eV , available below $2E_F$, with low losses. 
In this frequency range, the magnitude of the real part of wave number is $k_{||}=\text{Re}(K_{||})=10^8-10^9$/m, which is 1-2 orders of magnitude larger than the free space wave number $(=10^7)$/m \cite{Koppens2011,Abajo2014,Goncalves2016}. This feature of large wave number is interesting for strong  coupling of SP modes with emitters in the graphene environment. Furthermore the imaginary part $\kappa=\text{Im}(K_{||})$ is very small compared to real part, so SP graphene losses can be ignored below Fermi energy at sufficiently small plasmon propagation distances, a feature that we shall discuss further in later parts of this work. For energy below electron scattering rate $\hbar\tau^{-1}=0.01$ eV, shown clearly in Fig.\ref{fig:dispersion}b,c, the imaginary part of the complex wave number is larger or comparable to its real part and thus GSPs losses are significant in this low energy range. We therefore define our operating frequency as the range where losses are very low which spans the range $\hbar\tau^{-1}<\hbar\omega<2E_F$. For the case $E_F=0.5$ eV, this frequency range is about $0.01-0.8$ eV and increases for larger $E_F$. 
Finally we observe from Fig.\ref{fig:dispersion} that the effect of the dielectric layers asymmetry namely $\varepsilon_1=\varepsilon_2$ versus the case $\varepsilon_1\neq\varepsilon_2$ is not significant, we shall for the remaining of this paper focus only on the symmetric case 
$\varepsilon_1=\varepsilon_2=1$. 
A detailed anaylsis of the dielectric layers asymmetry has been carried out elsewhere \cite{Kamli2014}. 

\section{Graphene enhanced decay rates}

The graphene SP modes are localized interface eigenmodes that propagate parallel to interface in the $\hat{k}_{||}$ direction and decay away with increasing distance from interfaces at z=0. We saw in Fig.\ref{fig:dispersion} that graphene SP losses are very low $K_{||}\approx k_{||}$ for certain frequency range (e.g  0.01-0.8 eV for Fermi energy $E_F=0.5$ eV), then in this frequency range losses can be neglected, and the transverse quantized 2D SP fields that satisfy the Maxwell equations are constructed from the Fourier components of SP modes in the form \cite{Tame2013,Loudon2003,Matloob1995,Kamli2025}
 
\fla{
\hat{\textbf{E}}(\textbf{r},t)=& \frac{A}{4\pi^2}
\sum_{\lambda}\int d^2k_{||} \textbf{E}_{\lambda}(z,k_{||}) \hat{a}_{\lambda}(k_{||},t) e^{ik_{||}.r_{||}}
\nonumber
\\
+&H.C.,
\label{Electric-field}
}

\noindent
where $A$ is quantization area, $\lambda=p(s)$ refers to TM(TE) polarized  SP modes, and $\hat{a}_\lambda$ is the mode ($\lambda$) annihilation operator. The localized SP field amplitudes $\textbf{E}_\lambda(z,k_{||})=\mathcal{C}_\lambda(k_{||})\textbf{F}_\lambda(z,k_{||})$, and the classical mode structure functions $\textbf{F}_\lambda(z,k_{||})$ that satisfy wave equations and the boundary conditions, are given in the two half spaces, for TM (p) modes 
\fla{
{\textbf{F}_{p}}(z,k_{||})=&\text{u(z)} 
(\hat{\textbf{u}}_{||}+i\hat{\textbf{z}}\frac{ k_{||}}{\beta_1})e^{-\beta_1 z}+
\nonumber \\
& \text{u(-z)} 
(\hat{\textbf{u}}_{||}-i\hat{\textbf{z}}\frac{k_{||}}{\beta_2})e^{\beta_2 z},
\label{Exp-electric-field}
}     
and for TE (s) modes   
\fla{
&{\textbf{F}_{s}}(z,k_{||})=(\hat{\textbf{z}}\times\hat{\textbf{u}}_{||})[u(z) e^{-\beta_1 z}+u(-z)e^{\beta_2 z}]. 
\label{field_out}
}
 
The normalization factors $\mathcal{C}_{\lambda}$ specify the field amplitudes and are determined from the quantization procedure \cite{Tame2013,Ferreira2020,Kamli2025} that results in the  following expression  
\fla{
\mathcal{C}_{\lambda}(\omega)= \sqrt{\frac{\hbar\omega}{ \varepsilon_0 A L_{\lambda}(\omega)}}.
\label{Norm-factor}
}
For the TM polarized SP fields we have,
\fla{L_{p}(\omega)=D_{p}(\omega)+\frac{\omega^2}{c^2}S_{p}(\omega),
\label{LP}
}

\fla{
&\text{D}_{p}(\omega)= \frac{\varepsilon_1}{2\beta_1}\frac{|\beta_1|^2+|k_{||}|^2}{|\beta_1|^2} +
\frac{\varepsilon_2}{2\beta_2}\frac{|\beta_2|^2+|k_{||}|^2}{|\beta_2|^2},
\label{DP}
}

\fla{
&\text{S}_{p}(\omega)=
\frac{1}{2\beta_1} 
|\frac{\varepsilon_1}{\beta_1}|^2+\frac{1}{2\beta_1} 
|\frac{\varepsilon_2}{\beta_2}|^2.
\label{SP}
} 
Similarly for TE polarized SP modes we have 

\fla{L_{s}(\omega)=D_{s}(\omega)+\frac{c^2}{\omega^2}S_{s}(\omega),
\label{LS}
}

\fla{
\text{D}_{s}(\omega)=& \frac{\varepsilon_1}{2\beta_1} +\frac{\varepsilon_2}{2\beta_2},
\label{DW}
}
 
\fla{
&\text{S}_{s}(\omega)=\left[ \frac{|\beta_1|^2+|k_{||}|^2}{2\beta_1}+\frac{|\beta_2|^2+|k_{||}|^2}{2\beta_2}\right].
\label{SW}
}

The quantized SP field given by Eq.\eqref{Electric-field} can now be used to calculate the spontaneous emission rate due to surface plasmons near the graphene sheet. 
 We shall take the excited two level emitter as an excited two level atom, which relaxes to the ground state by spontaneous emission releasing a quantum of surface plasmon. The contribution of SP modes to the rate of spontaneous emission is evaluated in the Markov approximation from the Fermi golden rule \cite{Koppens2011,Loudon2003} subject to the condition \eqref{dispersion} for GSP to exist, and given  by the expression

\fla{
\Gamma_{sp}=\Gamma_{||}+\Gamma_z
\nonumber \\
\Gamma_{||}=3 \pi c \Gamma_0 \frac{d_{||}^2}{2 d^2} \left[ \frac{n_p^2}{k_{||}^p L_p}\frac{e^{-2\beta_1^p z}}{v_p}+\frac{n_s^2}{k_{||}^s L_s}\frac{e^{-2\beta_1^s z}}{v_s} \right]
\nonumber \\
\Gamma_{z}= 3 \pi c \Gamma_0 \frac{d_{z}^2}{d^2}   |\frac{k_{||}^p}{\beta_1^p}|^2 \frac{n_p^2}{k_{||}^p L_p}\frac{e^{-2\beta_1^p z}}{v_p}.
\nonumber
\\
\label{decay}
}

The SP group velocity $v_{(s,p)}=\partial\omega/\partial k_{||}$ is calculated from dispersion relation \eqref{dispersion} and shown in inset of Fig.\ref{fig:dispersion}a, and  $L_{s,p}$  
is the normalization length given in Eq.\eqref{Norm-factor}, and 

\fla{
\textit{n}_p (\omega)=\text{Re}\sqrt{\varepsilon_1-\left(\frac{\varepsilon_1}{\alpha f(\omega)}\right)^2\left(1+\frac{ \beta_1\varepsilon_2}{\beta_2\varepsilon_1}\right)^2},
\nonumber
 }
 
\fla{
\textit{n}_s (\omega)=\text{Re}\sqrt{\varepsilon_1-\left(\frac{\alpha f(\omega)}{2}\right)^2},
\label{nsp}
}
 
\noindent
and we have suppressed the temperature and Fermi energy dependence in the function $f(\omega)$. The atomic dipole emitter is located distance z above the interface with magnitude $d=(d_{||}^2+d_z^2)^{1/2}$ and $d_{||}$ and $d_z$ are the parallel and normal to interface components. 

\begin{figure}
    \centering
    \subfloat {(a){\includegraphics[width=8cm]{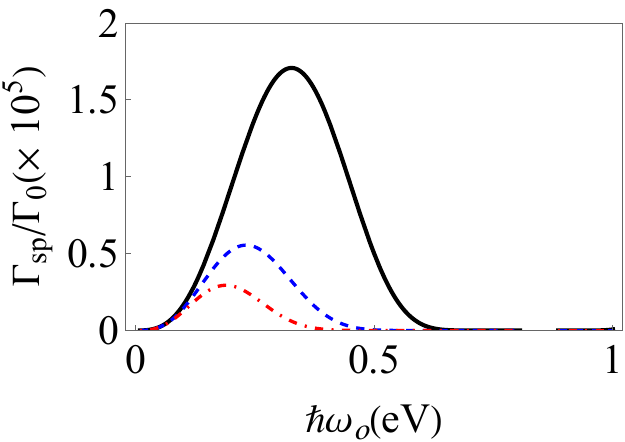}}}
    \qquad
    \subfloat{(b){\includegraphics[width=8cm]{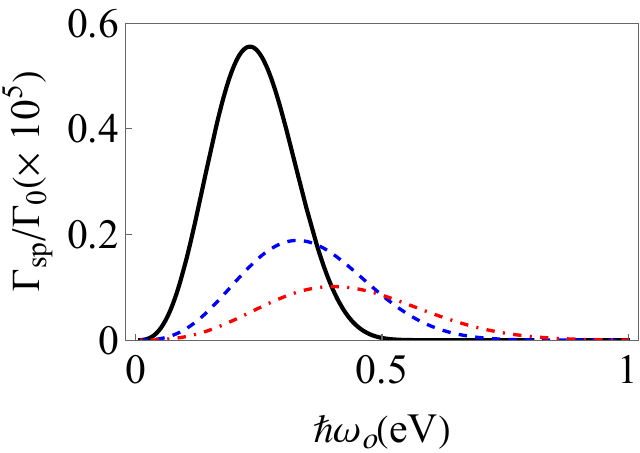}}}
    \caption{ (a) The SP Purcell enhancement factor $P_F=\Gamma_{sp}/\Gamma_0$ as a function of atomic transition frequency $\hbar\omega_0$ (eV) for dipole emitter placed $10$ nm (black solid line), $20$ nm (blue dashed line) and $30$ nm (red dotted line), for symmetric geometry $\varepsilon_1=\varepsilon_2=1$, $E_F=0.5$ eV and $\Omega=1.5\Gamma$. (b) The enhancement factor for different Fermi energies ; $E_F=0.5$ eV (black solid line), $E_F=1$ eV (blue dashed line), and $E_F=1.5$ eV (red dotted line) for $z=20$ nm, and $\Omega=1.5\Gamma$. }
    \label{fig:decay}
\end{figure}

The SP decay rate is given in terms of free space decay rate $\Gamma_0=\frac{d^2 \omega_0^3}{3\pi\varepsilon_0\hbar c^3}$, which is constant and for atomic transition in the visible region of spectrum has the value of the order  $\Gamma_0\approx10^8$/s \cite{Loudon2003}. The decay rate in free space ($\Gamma_0$) and due to SP modes ($\Gamma_{sp}$) is given as $\Gamma=\Gamma_0+\Gamma_{sp}$, which is a function of atomic position, z, and transition frequency $\omega_0$ etc. The SP decay rate contribution is shown in Fig.\ref{fig:decay} for the symmetric case $\varepsilon_1=\varepsilon_2=1$ for different dipole emitter locations (a) and different Fermi energies (b). The decay rate decreases with increasing distance from interface due to high field confinement, and also decreases as Fermi energy is increased. The decay rate attains a high peak for some transition frequency and suffers a dip near Fermi energy. This behavior of the decay rate can be seen with reference to wavenumber $k_{||}$ that we discussed before. The peaks are reduced and red-shifted to lower frequency as the atomic emitter recedes away from interface. They are also reduced but blue shifted to higher frequency as Fermi energy increases. The ratio  $P_F=\Gamma_{sp}/\Gamma_0$ gives the enhancement of spontaneous decay rate due to SP modes interacting with the dipole emitter near the graphene sheet and defines the enhancement factor. The SP decay rate is enhanced by 5-6 orders of magnitudes compared to free space decay rate $\Gamma_0$. The behavior of the decay rate and the large enhancement compare well with other works \cite{Koppens2011,Ferreira2020}. The decay rates depend on the location of the emitter above graphene sheet, on the Fermi energy etc as shown in Fig.\eqref{fig:decay}. Such rich parameters provide more options to control decay rates and other graphene-emitter dynamics. We shall need these results in the next section when we consider spectrum and coherence functions.
We note here that the quantization process outlined above is valid in the frequency range where losses are very low and can be ignored, it has also been used by many authors \cite{Tame2013,Loudon2003,Matloob1995,Ferreira2020}, and shown to be equivalent to more elaborate approaches as long as losses can be ignored which is the case throughout this paper.

\section{Resonance flourescence due to graphene surface plasmons}

 The dynamics of the interaction of a laser field driven two level emitter (two level atom), coupled to quantized SP field is described by the Hamiltonian of the form 
   
\fla{\hat{H}=\hat{H}_a+\hat{H}_f+\hat{H}_{int},
\label{11}
}
with

\fla{\hat{H}_a=&\frac{1}{2} \hbar\omega_0 \sigma_z ,
\\
\hat{H}_f=&\frac{1}{2} \int d^2k_{||}\hbar\omega(k_{||}) [\hat{a}^{\dagger}(k_{||},t)\hat{a}(k_{||},t)+H.C],
\\
\hat{H}_{int}=&-\frac{1}{2}\hbar\Omega \sigma_{+} e^{-i \omega_{L}t}-\sum_\lambda\int d^2k_{||}\hbar g_{k\lambda}(r) \sigma_{+} \hat{a}_{\lambda}(k_{||},t)
\nonumber
\\
+H.C.,
}
where  $\hat{H}_a$  is the atomic Hamiltonian with transition frequency $\omega_0$ and the two level atom operators are raising $\sigma_{+}=\ket{2}\bra{1}$, lowering $\sigma_{-}=\ket{1}\bra{2}$  operators and the inversion $\sigma_{z}=\frac{1}{2}\left(\ket{2}\bra{2}-\ket{1}\bra{1}\right)$,
$\hat{H}_f$  is the SP quantized field part with mode frequency $\omega(k_{||})$, 
and $\hat{H}_{int}$  is the dipole interaction Hamiltonian of the SP field with the two level atom, where the atom is positioned at $r=(r_{||0}=0,z>0)$ above interface in the upper medium $z>0$, and $\hbar$ is the reduced Planck constant. 
The atom - SP field coupling strength $g_{k\lambda}(r)=\textbf{d}\cdot \textbf{E}_\lambda(z,k_{||})e^{ik_{||}.r_{||}}/\hbar$ is a function of position $r=(r_{||},z)$, atomic transition frequency $\omega_0$ and other structure parameters. The classical driving field Rabi frequency is defined $\Omega=2\textbf{d}\cdot \textbf{E}_{L}/\hbar$
where $\textbf{E}_{L}$ is the driving field amplitude with frequency $\omega_L$.  The SP field amplitudes are as in Eqs \eqref{Electric-field} and $\textbf{d}$ is the dipole moment of the atomic transition.

The Heisenberg equation of motion $i\hbar\partial\hat{O}(t)/\partial t=[\hat{O}(t),\hat{H}]$ of operator $\hat{O}(t)$, relates the SP quantized field operator $\hat{a}(k_{||},t)$ and its conjugate to the atomic operators $\sigma_{\pm}(t)$ and we obtain the following expression for the positive frequency part of the field $\textbf{E}^{+}$ 

\fla{
\hat{\textbf{E}}^{+}(\textbf{r},t)=\frac{A}{4\pi^2}
\sum_{\lambda}\int d^2k_{||} |\mathcal{C}_{\lambda}(k_{||})|^2 \hat{\epsilon}_{\lambda} \textit{g}_{k\lambda} \times
\nonumber
\\
e^{ik_{||}.R_{||}} e^{-2\beta_{1}z} 
e^{-i\omega t} \int_{0}^{t} \sigma_{-}(t') e^{i\omega t'} dt'
\label{positive-field}
}
\noindent
where $R_{||}=\left|r_{||}-r_{||0}\right|$, $r_{||}$ is the field point, $g_{k\lambda}=\textbf{d}.\hat{\epsilon}_{\lambda}/\hbar$, $\hat{\epsilon}_{TM}=\hat{u}_{||}+i\hat{z}k_{||}/\beta_1$ for $z>0$ and $\hat{\epsilon}_{TM}=\hat{u}_{||}-i\hat{z}k_{||}/\beta_2$ for $z<0$, while for TE modes $\hat{\epsilon}_{TE}=(\hat{z}\times \hat{u}_{||})$. In Eq. \eqref{positive-field} we dropped a non-interacting part of the field since $<\hat{a}_{\lambda}(k_{||},0)>=0$.

To evaluate the angular integrations in Eq.\eqref{positive-field}, we define 
\fla{
\textbf{k}_{||}=k_{||}(\hat{x} \text{cos}\phi_k+\hat{y} \text{sin}\phi_k) ,
\nonumber
\\
\textbf{d}=d[(\hat{x}\text{cos}\phi_d+\hat{y}  \text{sin}\phi_d)\text{sin} \theta_d+\hat{z}\text{cos}\theta_d] 
\label{angles}
}
where $\phi_{d,k}$ are the angles the vectors $\textbf{d}$ and $\textbf{k}_{||}$ make with x-axis, and $\theta_{d}$ is the angle the dipole makes with z-axis. 
Now performing the angular integrations and assuming $\textbf{R}_{||}=R_{||}\hat{x}$ along x-direction we obtain after some manipulations, the following expression for the positive frequency part of SP field  
\fla{
\hat{\textbf{E}}^{+}(\textbf{r},t)=\frac{ d\omega_0^2}{4\pi\varepsilon_0 c} \mathcal{F}(\omega_0,z,R_{||},\theta)
&\sigma_{-}(t-\frac{R_{||}}{c}) 
\label{SP-field}
}

The function $\mathcal{F}(\omega_0,z,R_{||},\theta)$ describes the field profile and represents the modification of the graphene sheet to the surface plasmon emitted field. It depends on the emitter and graphene parameters and contains all information about the field profile and is given by the closed form exact expression
\fla{
\mathcal{F}(\omega_0,z,R_{||},\theta)=\pi \text{sin}\theta_d [\frac{ n_p \mathcal{F}_p(\omega_0,R_{||},\theta) }{L_p v_g^p} e^{-2\beta_1^p z}+
\nonumber
\\
\frac{ n_s \mathcal{F}_s(\omega_0,R_{||},\theta)} {L_s v_g^s} 
e^{-2\beta_1^s z}] e^{-i\omega_0 R_{||}/c} ,
\label{F-functions}
}

\begin{figure}
\centering
    \subfloat{(a){\includegraphics[width=8cm]{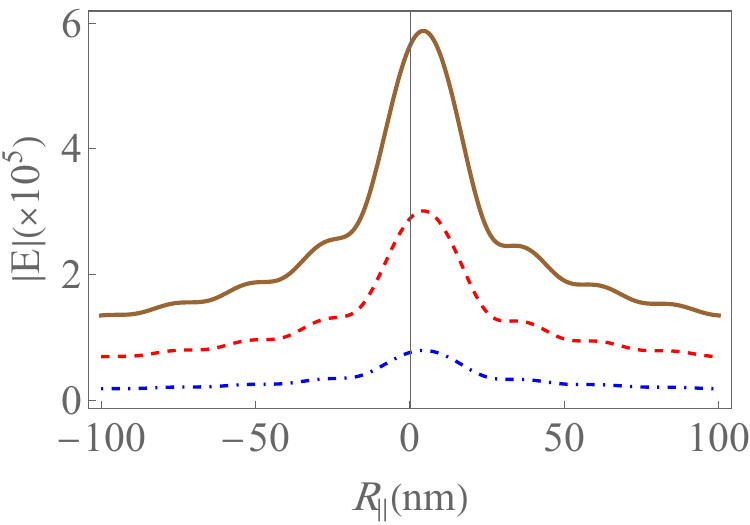}}}
    \qquad
    \subfloat{(b){\includegraphics[width=8cm]{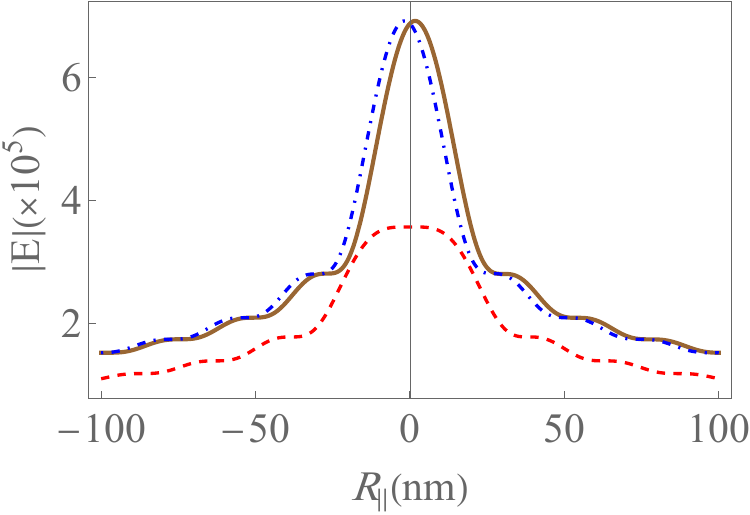}}}
    \caption{The field profile function $|E|$ as a function of separation $R_{||}$ nm for (a) different graphene-emitter separations $z=5$ nm brown solid line, $z=10$ nm dotted red line and $z=20$ nm dashed blue line. (b) different dipole emitter orientations $\theta_d=\pi/10$ ( upper blue dashed line), $\theta_d=\pi/2$ ( brown solid line), and $\theta_d=0.9\pi$ ( red dotted line), $\phi=\pi/4$. These figures show oscillatory behavior of field profile indicating strong SP-atom interaction. }
    
    \label{fig:profile}
\end{figure}

\fla{
\mathcal{F}_p=
2\hat{x}\left(\text{cos}\phi_d \left(\frac{J_1(u)}{u}-J_2(u)\right )-J_1(u) \frac{k_{||}}{\beta_1} \text{cot}\theta_d \right)+
\nonumber
\\
2 \hat{y} \text{sin}\phi_d \frac{J_1(u)}{u} -2
\hat{z} \frac{k_{||}}{\beta_1}\left(\text{cos}\phi_d J_1(u)+J_0(u) \text{cot}\theta_d \frac{k_{||}}{\beta_1} \right)
}
\fla{
\mathcal{F}_s=2\hat{x} \text{cos}\phi_d \frac{J_1(u)}{u}+ 2\hat{y} \text{sin}\phi_d \left(\frac{J_1(u)}{u}-J_2(u)\right)  
\label{fig:p}
}

\noindent
where $\theta$ refers to dipole orientation angles $\theta_d$, and $\phi_d$, and $J_n(u)$ are the Bessel functions of order n and $u=k_{||}R_{||}$. The above expression is exact and valid for near and far fields. We shall retain here the exact field profile and discuss the far field case in the Appendix. The emitted SP field profile given by the quantity $|E|=(d\omega_0^2/4\pi\varepsilon_0c)\mathcal{F}(\omega_0,z,R_{||},\theta)$
is shown in Fig.\eqref{fig:profile} as functions of graphene sheet emitter separations $z$ and $R_{||}$, the dipole emitter orientations $\theta_d, \phi_d$ etc. Fig.\eqref{fig:profile}(a) shows the field profile as a function of in-plane separation $R_{||}$ (in nm) for different graphene sheet emitter separation $z$ normal to graphene interface, while Fig.\eqref{fig:profile}(b) shows the field profile for varied dipole orientations ($0<\theta_d<\pi$). Near the interface the field is highly enhanced due to confinement and decreases exponentially with distance from interface. In the in-plane direction parallel to graphene sheet, the field also decreases with distance from the emitter and shows oscillatory behavior due to modification of function $\mathcal{F}$ which depends on the dipole emitter and graphene structure parameters, a behavior indicative of strong SP-atom coupling. 
\noindent

The exact field expression  \eqref{SP-field} will be used to explore the spectral and statistical properties of the emitted field. The spectrum function $S(r,\omega_0)$ for the transition $\omega_0$ measured at $r=(R_{||},z)$ is defined as the Fourier transform of the SP field correlation;

\fla{
\text{S}(r,\omega_0)= \frac{1}{\pi} Re\int d\tau e^{i\omega_0 \tau}  <E^{-}(t)E^{+}(t+\tau)>
}
where from \eqref{SP-field} ( neglecting $R_{||}/c$ term) we have 
\fla{
<E^{-}(t)E^{+}(t+\tau)>= I_s<\sigma_{+}(t)\sigma_{-}(t+\tau)>
\label{correlation}
}
and the field intensity is  
\fla{
\textit{I}_s(r,\omega_0,\theta)=\left(\frac{d \omega_0^2 }{4\pi\varepsilon_0 c}\right)^2 \left|\mathcal{F}(\omega_0,z,R_{||},\theta)\right|^2. 
\label{intensity}
}
 Now to proceed with evaluation of the two time operator correlations in Eq. \eqref{correlation} we follow standard resonance fluorescence calculations \cite{Mandel1995,Scully1997}, write down master equation, use well known solutions of Bloch equations and resort to quantum regression formula to evaluate the two time correlations, one obtains the expression  

\fla{
\text{S}(r,\omega_0,\theta)= \frac{I_s(r,\omega_0,\theta)}{2\pi}   \frac{\Omega^2}{\Gamma^2+2\Omega^2} \times
\nonumber
}
\fla{
\left[\frac{\Gamma/2}{(\frac{\Gamma}{2})^2+\Delta^2}+\frac{\alpha_{+}/2}{(\frac{3\Gamma}{4})^2+(\Delta+\mu)^2}+\frac{\alpha_{-}/2}{(\frac{3\Gamma}{4})^2+(\Delta-\mu)^2} \right].
\label{spectrum}
}

In this expression we dropped an elastic scattering part proportional to $\delta({\Delta})$ function where $\Delta=\omega_0-\omega$, $\alpha_{\pm}=(3\Gamma/4)P{\pm}(\omega{\pm}\mu-\omega_{0})Q$, $P=(2\Omega^2-\Gamma^2)/(2\Omega^2+\Gamma^2)$, $Q= (\Gamma/4\mu)(10 \Omega^2-\Gamma^2)/(2\Omega^2+\Gamma^2)$, and $\mu=\sqrt{\Omega^2-\Gamma^2/16}$.

The quantity $\Gamma$ is the spontaneous decay rate $\Gamma=\Gamma_0+\Gamma_{sp}$, where the SP contribution is calculated from the equation of motion of the excitation level $d<\sigma_+(t)\sigma_-(t)>/dt$ which results in Eq.\eqref{decay} for decay rate in the Markov approximation. The decay rate is a function of atom position, transition frequency, dipole orientations etc which we suppress for notational convenience. 

\begin{figure}
    \centering
    \subfloat{{(a)}{\includegraphics[width=8cm]{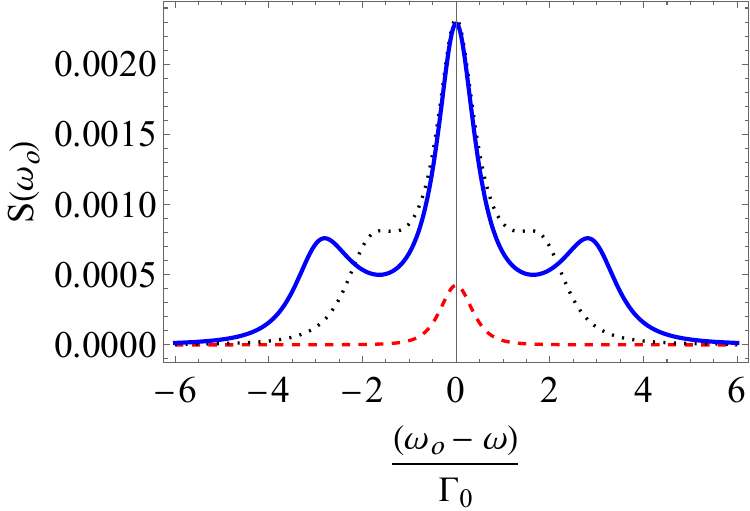}}}
    \qquad
    \subfloat{{(b)}{\includegraphics[width=8cm]{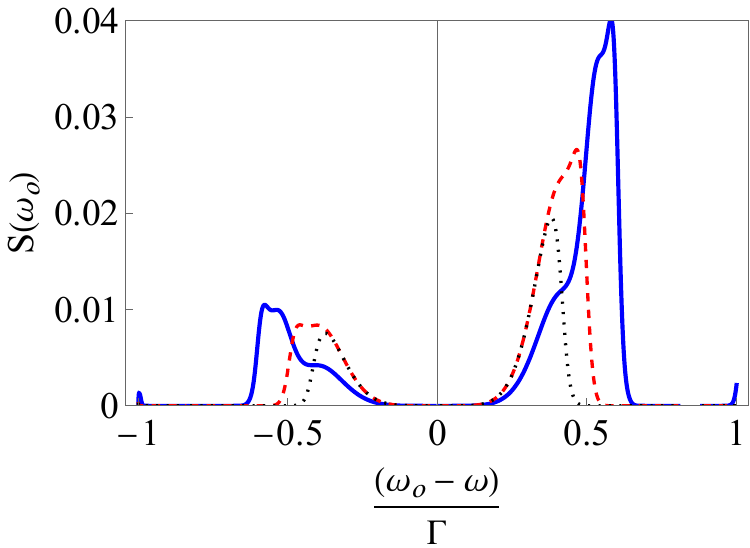}}}
    \qquad
    \subfloat{{(c)}{\includegraphics[width=8cm]{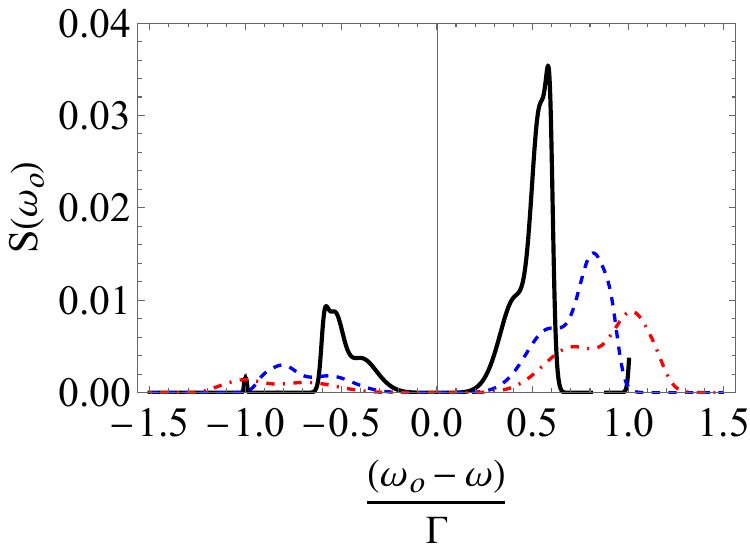}}}

    \caption{ (a). The spectrum function as a function of atomic transition frequency $\omega_0$ in free space for Rabi frequency $\Omega=0.2\Gamma_0$ (red dashed line), $\Omega=1\Gamma_0$ (black dotted line) and $\Omega=1.5\Gamma_0$ (blue solid line). (b) The spectrum for an atom near the graphene sheet $z=20$ nm (blue solid), $z=35$ nm (red dotted), $z=50$ nm (black dashed), for Rabi frequency $\Omega=1.5 \Gamma$.  (c) The spectrum for an atom $z=20$nm above the graphene sheet and Rabi frequency  $\Omega=1.5\Gamma$ for different Fermi energy $E_F=0.5$ eV (black solid line) ; $E_F=1$ eV (blue dashed line), and $E_F=1.5$ eV (red dotted line) .}
    \label{fig:spectrum}
\end{figure}

The behavior of the spectrum function is much affected by the behavior of the SP decay rates near the graphene sheet and by the modification of the intensity function $I_s$ given in expression \eqref{intensity} in terms of the functions $\mathcal{F}$.  
The spectrum is shown in Fig.\ref{fig:spectrum} as a function of atomic transition frequency $\omega_0$ in free space case (a) for Rabi frequency $\Omega=0.2\Gamma_0$ (red dashed line), $ \Omega=1\Gamma_0$ (black dotted line) and $\Omega=1.5\Gamma_0$ (blue solid line), showing a central peak and two side bands symmetric peaks that constitute Mollow triplet \cite{Mollow1969}. Fig.\ref{fig:spectrum}(b) shows the modification of spectrum for the emitter above the graphene sheet at $z=20$ nm (blue solid), $z=35$ nm (red dotted), $z=50$ nm (black dashed), for Rabi frequency $\Omega=1.5 \Gamma$. The spectrum peaks are higher when the atomic emitter is near the graphene interface since the interaction with SP excitations is stronger due to field confinement. As the emitter is farther away from interface interaction diminishes and the spectrum peaks are reduced. Here we observe the two sideband peaks and suppressed central peak. The other two sharp peaks on the far right and left of Fig.\ref{fig:spectrum}b are due to the behavior of the quantities $n_p,n_s$ that appear in the function $\mathcal{F}$ that are related to the wave number $k_{||}$ as we discussed before. The positions of these sharp peaks shift with dipole emitter location near interface. Fig.\ref{fig:spectrum}c shows the spectrum for the atom $z=20$ nm above the graphene sheet and different Fermi energies; $E_F=0.5$ eV (black solid line), $E_F=1$ eV (blue dashed line), and $E_F=1.5$ eV (red dotted line). An important feature here in Figs.\ref{fig:spectrum}(b,c) is the high suppression of the central peak in the presence of the graphene sheet. This is attributed to the high enhancement of the decay rate (see Fig.\ref{fig:decay}) that enters into the expression for the spectrum, since the decay rate is enhanced and for Rabi frequency comparable to linewidth, the spectrum behaves like $\Gamma^{-1}$. So by positioning the dipole emitter near the graphene sheet and/or changing the Fermi energy of the graphene sheet one can control and modify the behavior of the spectrum. Control of two level emitter (or qbit) and its spectrum is important in various information and communication processing and technological applications.

\begin{figure}
\centering
    \subfloat {(a){\includegraphics[width=8cm]{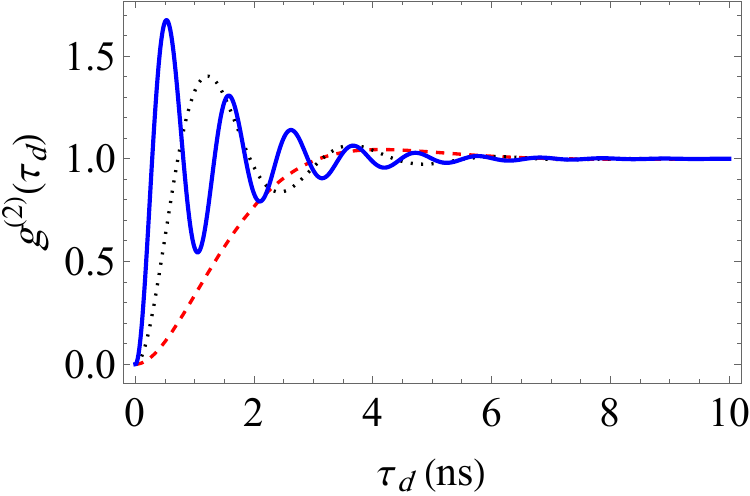}}}
    \qquad
    \subfloat{(b){\includegraphics[width=8cm]{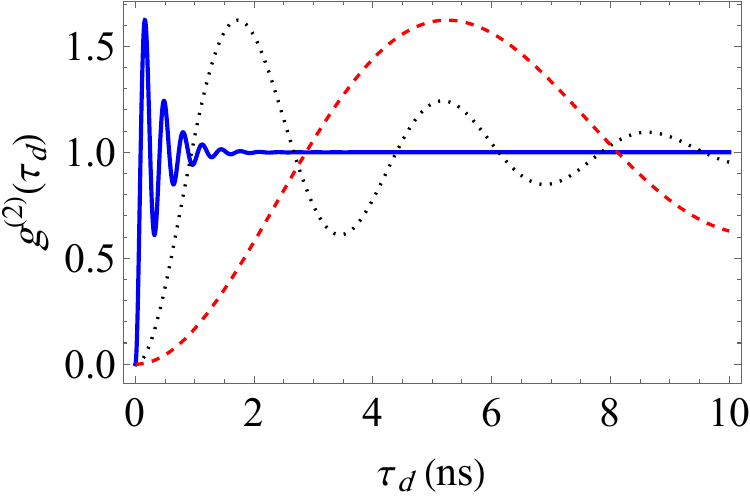}}}
    \qquad
    \subfloat{(c){\includegraphics[width=8cm]{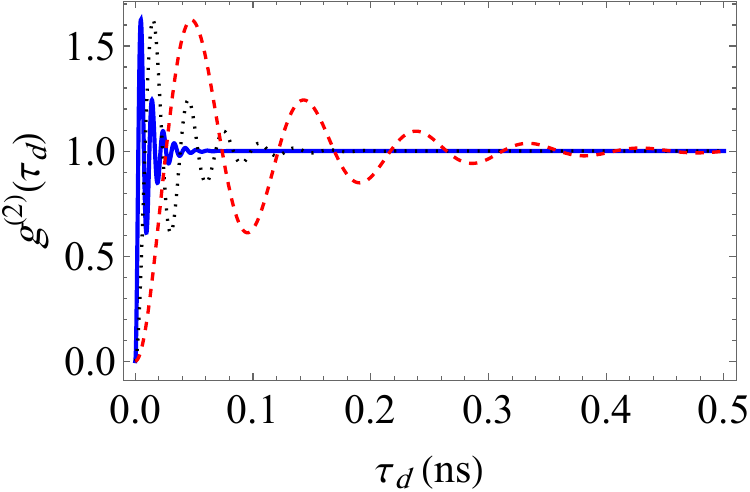}}}
    \caption{ (a). The second order coherence function  $g^{(2)}(\tau_d)$ as a function of time delay $\tau_d$ (ns) in free space for atomic transition frequency $\hbar\omega_0=0.5 $eV and Rabi frequency $\Omega=0.2\Gamma$ (red dashed line), $\Omega=1\Gamma$ (black dotted line) and $\Omega=1.5 \Gamma$ (blue solid line). (b) The coherence function $g^{(2)}(\tau_d)$ as a function of time delay for emitter near graphene sheet at distance $z=40$ nm (blue solid), $z=50$ nm (black dashed) and $z=60$ nm (red dotted) for Fermi energy $E_F=0.5$ eV and Rabi frequency $\Omega=1.5 \Gamma$.(c) as in (b) but for $E_F=1$ eV.}
    \label{fig:coherence1}
\end{figure}

\begin{figure}
\centering
    \subfloat{(a){\includegraphics[width=8cm]{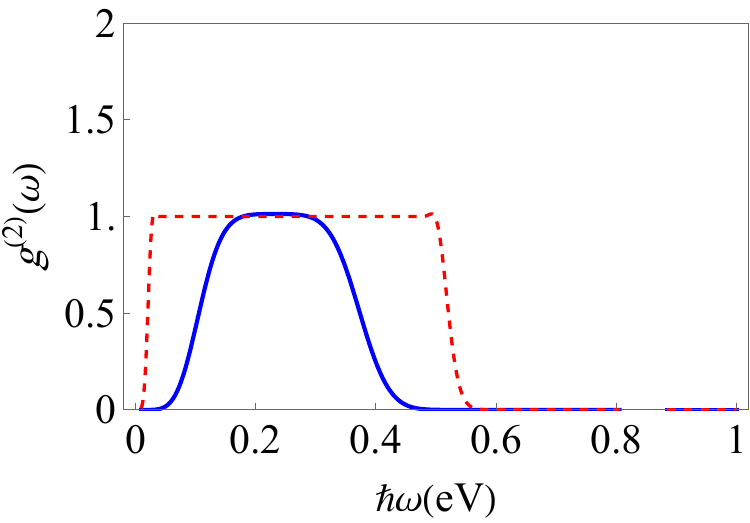}}}
    \qquad
    \subfloat{(b){\includegraphics[width=8cm]{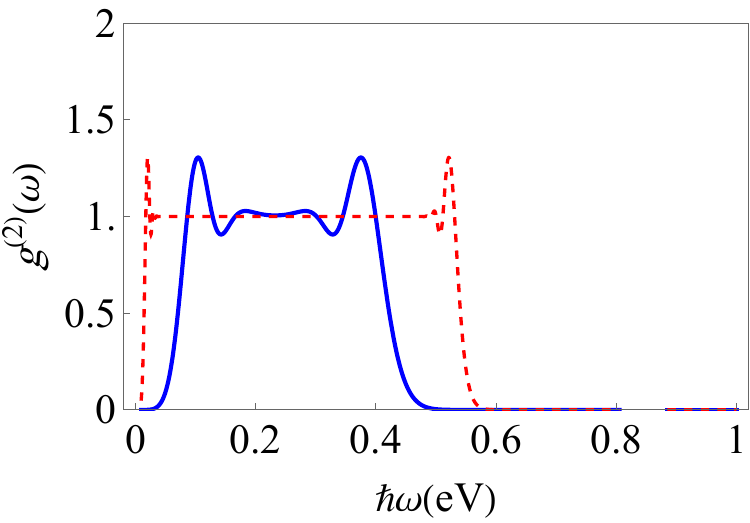}}}
    \qquad
    \subfloat{(c){\includegraphics[width=8cm]{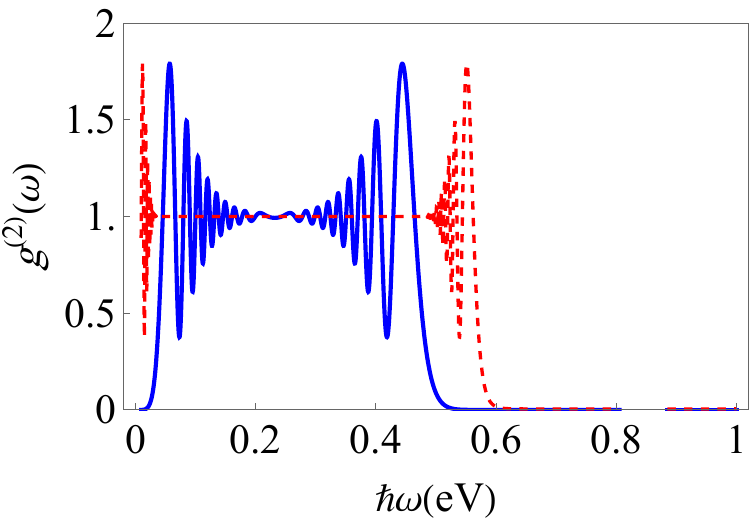}}}
    
    \caption{ The coherence function $g^{(2)}(\omega)$ as a function of SP mode frequency for Rabi frequency $\Omega=0.5 \Gamma$ (a), $\Omega=2 \Gamma$ (b) and $\Omega=10 \Gamma$ (c). The red dashed line is $\tau_d=0.1$ ns and blue solid line is $\tau_d=0.001$ ns, $z=20$ nm.}
    \label{fig:coherence2}
\end{figure}

\section{Coherence function and antibunching}

It is well established that the radiation emitted in resonance fluorescence shows non-classical features namely anti-bunching effect. This is described by the second order coherence function $g^{(2)}(\tau_d)$ that measures the intensity-intensity correlation, determined from the solutions to optical Bloch equations and given by the expression \cite{Scully1997}

\fla{
\textit{g}^{(2)}(\tau_d)=1-e^{-\frac{3}{4}\Gamma\tau_d} \left[ \text{cos}(\mu\tau_d)+\frac{3\Gamma}{4\mu}\text{sin}(\mu\tau_d)\right]
\label{coherence}
}

From this equation we see that for zero time delay $\tau_d=0$, the function $g^{(2)}(0)=0$, while for long time delay $\tau_d\rightarrow\infty$ it saturates to the value $g^{(2)}\rightarrow1$ showing non-classical character $g^{(2)}(\tau_d)>g^{(2)}(0)$ and hence antibunching effect discussed long ago \cite{Carmichael2002} and observed experimentally  \cite{Kimble1977}. This effect indicates that the SPs emitted during resonance fluorescence arrive at the detection device not as bunches but as individual photons separated by time lapse $\tau_d$. The intermediate time behavior of $g^{(2)}(\tau_d)$ depends on the size of Rabi frequency relative to spontaneous emission rate $\Gamma$, which is a function of emitter position, transition frequency, graphene sheet parameters. In free space the decay rate is constant ($\Gamma=\Gamma_0$), and for weak Rabi frequency $\Omega<\Gamma$ the function $g^{(2)}(\tau_d)$ starts from zero time delay and increases monotonically with time, while for large Rabi frequency $\Omega>\Gamma$ the coherence function oscillates and tends to 1 after long time delay as is shown in Fig.\ref{fig:coherence1}a. In the presence of graphene sheet, the spontaneous rate $\Gamma$ is no longer constant, but modified and enhanced as in Fig.\ref{fig:decay} and thus the coherence function is expected to be modified accordingly. This modification is shown in Fig.\ref{fig:coherence1}b for dipole emitter at different distances z from graphene sheet: $z=40$ nm (blue solid), $z=50$ nm (black dashed) and $z=60$ nm (red dotted) for Rabi frequency $\Omega=1.5 \Gamma$ and Fermi energies $E_F=0.5$ eV (b) and $E_F=1$ eV in (c). In these graphs the coherence function oscillates with time due to strong Rabi frequency in a manner that depends on the distance of the emitter from graphene sheet and on the Fermi energy. Close to the interface the decay rate is enhanced and thus oscillations die out quickly before reaching value of 1, while away from interface the decay rate is smaller and hence the oscillations persist for longer times before saturate to 1.  The effect of increasing the Fermi energy is to shift the curves into shorter time delays as shown in Figs. (b) and (c), which provides additional dynamic control on how the emitted surface plasmons are correlated in the presence of the graphene sheet, a feature that could be of potential interest in quantum photonics. 
In Figs.   
\ref{fig:coherence2} we plot $g^{(2)}(\tau_d,\omega)$ as a function of SP mode energy $\hbar\omega$(eV) for varying time delays and different Rabi frequencies. The behavior of the coherence function with SP energy is explained by reference to the two level emitter  modified decay rate, which has a nearly bell like shape shown in Fig.\ref{fig:decay} and also by the strength of Rabi frequency as discussed in Fig.\ref{fig:coherence1}. For low SP excitation frequency the decay rates are highly suppressed, which leads to $g^{(2)}\rightarrow0$. In the intermediate regions of SP frequency the decay rates are highly enhanced leading to $g^{(2)}\rightarrow1$. At higher SP frequency such that $Re[n_{p,s}]=0$, the decay rate is highly reduced again and coherence oscillations revive, where the coherence settles at $g^{(2)}(\tau,\omega_r)=1$ for frequency $\omega_r$ is the root of the function $Re[n_{p,s}(\omega_r)]=0$ which is a strongly graphene conductivity dependent quantity, and then die again as the frequency approaches $2E_F$. This feature is interesting since it can provide another means to control resonance spectrum and influence the behavior of the coherence by using graphene electric gating and doping to manipulate its conductivity and provide additional control.

\section{Conclusion}

We investigated the emission of graphene surface plasmons in two level emitter resonance fluorescence near a 2D graphene sheet. We derived an exact closed form expression for the emitted SP field valid in the near and far regions. We calculated the SP decay rates and the modified spectrum and coherence functions due to graphene 2D sheet structure and obtained huge enhancement compared to free space decay linewidth. Calculations were performed for different parameters including  locations of emitter from interface, dipole orientation angles and  for different Fermi energies etc. 
The GSP field excitation was found to show high confinement and strongly depend on the distance between the emitter and the graphene surface and on the dipole orientation angles. The analysis showed that even with a small increase in the distance $\delta z$ from interface, there is a sharp decrease (a factor of $1/e^{(\beta_1\delta z)}$) in the amplitude of the radiated field. It is interesting that at small distances parallel to the graphene sheet, the field does not decrease so much with distance from the emitter. This decrease is accompanied by oscillations of the SP amplitude indicative of strong interactions of emitter-GPS which can be attributed to the highly confined SP fields near graphene interface as discussed earlier in the text. 
We have performed simulations for the spectrum function $S(\omega_0,z,\theta)$ and second order coherence function 
$g^{(2)}(\tau_d,z)$ for various structure parameters. The analysis showed rich and possible dynamic control parameters most notably the Fermi energy, that can be used to tailor spectrum and second order coherence functions for practical applications that may be useful for integration with quantum photonics, without changing the size of the structure environment e.g the dielectric layers surrounding the graphene sheet.   
Another important feature of the excited local field and its resonance fluorescence dynamics is the strong dependence on the orientation angles of the dipole moment of the emitter. It is interesting to formulate here the role of the dipole emitter orientations and the possibility to develop ways to control resonance fluorescence and coherence functions based on the emitter orientations.
The amalgamation of the two level emitter dynamics and the graphene structure properties with possible dynamic control of conductivity presents a rich set of control tools that can be utilized to engineer and manipulate the spectrum and statistics in the emitted SP fields, a feature that is of great importance in modern information processing and communications and potentially useful for various photonics applications.

\section{ACKNOWLEDGMENTS}
The authors gratefully acknowledge the funding of the Deanship of Graduate Studies and Scientific Research, Jazan University, Saudi Arabia, through Project number: JU- 202503186 -DGSSR-RP-2025.

\section{Appendix} 

The exact form of the vector function $\mathcal{F}(\omega_0,z,R_{||},\theta)$ is 
\fla{
\mathcal{F}(\omega_0,z,R_{||},\theta)=\pi \text{sin}\theta_{d}[\frac{n_p \mathcal{F}_p(\omega_0,R_{||},\theta) }{L_p v_g^p} e^{-\beta_1^p(z+z_0)}+
\nonumber
\\
\frac{n_s \mathcal{F}_s(\omega_0,R_{||},\theta)} {L_s v_g^s} 
e^{-\beta_1^s(z+z_0)}] e^{-i\omega_0 R_{||}/c} ,
\label{F-functions}
}

\fla{
\mathcal{F}_p(\omega_0,R_{||},\theta)=2\hat{x}\left(\text{cos}\phi_d \left(\frac{J_1(u)}{u}-J_2(u)\right )-J_1(u) \frac{k_{||}}{\beta_1} \text{cot}\theta_d \right)+
\\
2 \hat{y} \text{sin}\phi_d \frac{J_1(u)}{u} -2
\hat{z} \frac{k_{||}}{\beta_1}\left(\text{cos}\phi_d J_1(u)+J_0(u) \text{cot}\theta_d \frac{k_{||}}{\beta_1} \right)
}
\fla{
\mathcal{F}_s(\omega_0,R_{||},\theta)=2\hat{x} \text{cos}\phi_d \frac{J_1(u)}{u}+ 2\hat{y} \text{sin}\phi_d \left(\frac{J_1(u)}{u}-J_2(u)\right)  
\label{Fp}
}

\noindent
where $J_n(u)$ is the Bessel function of order n and $u=k_{||}R_{||}$. The angle $\theta$ refers to $\theta_d$ or $\phi_d$. 
The far field limit of the functions $\mathcal{F}$ is obtained by taking the asymptotic form   $u=k_{||}R_{||}\rightarrow\infty$ of Bessel functions $J_n(u)$, 
\fla{
\textit{J}_n(u)= \sqrt{\frac{2}{\pi u}} \text{cos}\left[u-\frac{\pi}{2}(n+\frac{1}{2}\right]
\label{Bessel}
}
and gives 
\fla{
\textit{J}_0(u)= \frac{1}{\sqrt{\pi u}} (\text{sin}u+\text{cos}u)=-\textit{J}_2(u) , 
\nonumber
\\
\textit{J}_1(u)= \frac{1}{\sqrt{\pi u}} (\text{sin}u-\text{cos}u).
\nonumber
\label{Bessel}
}
Then the far field functions $\mathcal{F}_{s,p}$ reduce to 

\fla{
\mathcal{F}_{p}(R_{||}\rightarrow\infty)=\frac{1}{\sqrt{u}} \textbf{G}_{p} , \mathcal{F}_{s}(R_{||}\rightarrow\infty)=\frac{1}{\sqrt{u}} \textbf{G}_{s}
}
and the far field vector function $\mathcal{F}$ becomes 

\fla{
\mathcal{F}_{far}(\omega_0,z,R_{||},\theta)=\frac{1}{\sqrt{u}}\textbf{G}(\omega_0,z,R_{||},\theta)
\label{F-far}
\\
\textbf{G}(\omega_0,z,R_{||},\theta)=\pi \text{sin}\theta_{d} [\frac{n_p \textbf{G}_p(\omega_0,R_{||},\theta) }{L_p v_g^p} e^{-\beta_1^p(z+z_0)}+
\nonumber
\\
\frac{n_s \textbf{G}_s(\omega_0,R_{||},\theta)} {L_s v_g^s} 
e^{-\beta_1^s(z+z_0)}] e^{-i\omega_0 R_{||}/c} ,
\label{G-functions}
}

\fla{
\textbf{G}_{p}(\omega_0,R_{||},\theta)=\frac{2}{\sqrt{\pi}}\times
\nonumber
\\
\nonumber
\hat{x}
[\text{cos}\phi_d (\text{sin} u+\text{cos} u)- 
\frac{k_{||}}{\beta_1}\text{cot}\theta_d (\text{sin} u-\text{cos}u)]
\label{Fp-far}
\\
\nonumber
-\hat{z}\frac{k_{||}}{\beta_1}[\text{cos}\phi_d (\text{sin} u-\text{cos} u)+ 
\frac{k_{||}}{\beta_1}\text{cot}\theta_{d} (\text{sin} u+\text{cos}u)] ,
}

\fla{
\textbf{G}_s(\omega_0,R_{||},\theta)= \hat{y} \frac{2}{\sqrt{\pi}} \text{sin} \phi_d (\text{sin}u+\text{cos} u).
\nonumber
}

Using Eqs.\eqref{F-far} and \eqref{G-functions} into the the full function $\mathcal{F}$ and inserting into the expression \eqref{SP-field} we obtain for the SP far field  

\fla{
\hat{\textbf{E}}_{far}^{+}(\textbf{r},t)=\frac{ d\omega_0^2}{4\pi\varepsilon_0 \sqrt{R_{||}} c^2} \textbf{G}(\omega_0,R_{||},z)
&\sigma_{-}(t-\frac{R_{||}}{c}).
}
The total radiated power is calculated over a circular ring of raduis $R_{||}$ and centered around the atom as $\approx2\pi R_{||} |E|^2$.

\bibliography{main}

\end{document}